\documentclass[aps,prc,twocolumn,nofootinbib,amsmath,showpacs,superscriptaddress,floatfix]{revtex4}
\usepackage{graphicx,epsfig,longtable,lscape,rotating}
\usepackage{mathptmx}
\usepackage[light,first]{draftcopy} 
\usepackage{epstopdf}  
\usepackage[pdftex]{hyperref}
\usepackage{wasysym}

\newcommand{\beqy}{\begin{eqnarray}}
\newcommand{\eeqy}{\end{eqnarray}}
\newcommand{\bmlet}{\begin{subequations}}
\newcommand{\emlet}{\end{subequations}}
\newcounter{saveeqn}

\def\gsimeq{\,\,\raise0.14em\hbox{$>$}\kern-0.76em\lower0.28em\hbox  
{$\sim$}\,\,}  
\def\lsimeq{\,\,\raise0.14em\hbox{$<$}\kern-0.76em\lower0.28em\hbox  
{$\sim$}\,\,}  

\begin{document}

\title{Experimentally constrained ($p,\gamma$)$^{89}$Y and ($n,\gamma$)$^{89}$Y reaction rates \\
relevant to the $p$-process nucleosynthesis}

\author{A.~C.~Larsen}
\email{a.c.larsen@fys.uio.no}
\affiliation{Department of Physics, University of Oslo, N-0316 Oslo, Norway}
\author{M.~Guttormsen}
\email{magne.guttormsen@fys.uio.no}
\affiliation{Department of Physics, University of Oslo, N-0316 Oslo, Norway}
\author{R.~Schwengner}
\email{r.schwengner@hzdr.de}
\affiliation{Helmholtz-Zentrum Dresden-Rossendorf, 01328 Dresden, Germany}
\author{D.~L.~Bleuel}
\affiliation{Lawrence Livermore National Laboratory, Livermore, California 94551, USA}
\author{S.~Goriely}
\affiliation{Institut d'Astronomie et d'Astrophysique, Universite Libre de Bruxelles, Brussels, Belgium}
\author{S.~Harissopulos}
\affiliation{Institute of Nuclear Physics, NCSR “Demokritos”, Athens, Greece}
\author{F.~L.~Bello~Garrote}
\affiliation{Department of Physics, University of Oslo, N-0316 Oslo, Norway}
\author{Y.~Byun}
\affiliation{Department of Physics and Astronomy, Ohio University, Athens, Ohio 45701, USA}
\author{T.~K.~Eriksen}
\affiliation{Department of Physics, University of Oslo, N-0316 Oslo, Norway}
\author{F.~Giacoppo}
\affiliation{Department of Physics, University of Oslo, N-0316 Oslo, Norway}
\author{A.~G{\"o}rgen}
\affiliation{Department of Physics, University of Oslo, N-0316 Oslo, Norway}
\author{T.~W.~Hagen}
\affiliation{Department of Physics, University of Oslo, N-0316 Oslo, Norway}
\author{M.~Klintefjord}
\affiliation{Department of Physics, University of Oslo, N-0316 Oslo, Norway}
\author{T.~Renstr{\o}m}
\affiliation{Department of Physics, University of Oslo, N-0316 Oslo, Norway}
\author{S.~J.~Rose}
\affiliation{Department of Physics, University of Oslo, N-0316 Oslo, Norway}
\author{E.~Sahin}
\affiliation{Department of Physics, University of Oslo, N-0316 Oslo, Norway}
\author{S.~Siem}
\affiliation{Department of Physics, University of Oslo, N-0316 Oslo, Norway}
\author{T.~G.~Tornyi}
\affiliation{Department of Physics, University of Oslo, N-0316 Oslo, Norway}
\author{G.~M.~Tveten}
\affiliation{Department of Physics, University of Oslo, N-0316 Oslo, Norway}
\author{A.~V.~Voinov}
\affiliation{Department of Physics and Astronomy, Ohio University, Athens, Ohio 45701, USA}
\author{M.~Wiedeking}
\affiliation{iThemba LABS, P.O.  Box 722, 7129 Somerset West, South Africa}

\date{\today}

\begin{abstract}
The nuclear level density and the $\gamma$-ray strength function have been extracted for $^{89}$Y,
using the Oslo Method on $^{89}$Y($p,p' \gamma$)$^{89}$Y
coincidence data. The $\gamma$-ray strength function displays a low-energy enhancement
consistent with previous observations in this mass region ($^{93-98}$Mo). Shell-model calculations give support that
the observed enhancement is due to strong, low-energy $M1$ transitions at high excitation energies. 

The data were further
used as input for calculations of the $^{88}$Sr($p,\gamma$)$^{89}$Y and $^{88}$Y($n,\gamma$)$^{89}$Y
cross sections with the TALYS reaction code. 
Comparison with cross-section data, where available, as well as with values from the BRUSLIB library,
shows a satisfying agreement.
\end{abstract}


\maketitle

\section{Introduction}
\label{sec:int}
The quest for a detailed understanding of  heavy-element nucleosynthesis is very intriguing,
calling for an extensive knowledge of nuclear properties at the relevant astrophysical energies.
The astrophysical aspects are equally challenging; the most critical issue is to
identify the correct astrophysical sites and conditions
in which the heavy-element nucleosynthesis
operates~\cite{arnould2007,arnould2003,iliadis2007,rauscher2013,sneden2008}. 

There are three main processes responsible for creating elements heavier than iron~\cite{burbidge1957}: 
the slow neutron-capture ($s$-) process, the rapid neutron-capture ($r$-) process, and the $p$-process, 
which in fact includes a variety of nuclear reactions such as photodisintegration of already created 
$s$- and $r$-nuclides through e.g. ($\gamma,p$), ($\gamma,n$), and ($\gamma,\alpha$) reactions 
as well as their inverse capture reactions. To this end, only the $s$-process astrophysical sites are more or less clearly
identified (low-mass asymptotic giant branch stars, $M \lesssim 8 M_{\odot}$, and massive stars, $M > 8 M_{\odot}$;
see e.g. Ref~\cite{iliadis2007}).
The $r$-process astrophysical site is still a mystery, suggestions include the neutrino-driven
wind following a core-collapse supernova, and neutron-star mergers~\cite{arnould2007,just15}. Also the $p$-process site remains 
rather elusive~\cite{arnould2003,rauscher2013,travaglio2011}; however, it is clear that conditions with temperatures reaching between
2 to 3 billions degrees must 
be reached for the $p$-process to take place. The deep O-Ne layers of a massive star prior to or in its supernova
phase remains the most popular suggestion, though some species, in particular the $p$-isotopes of Mo and Ru, remain
underproduced in the present simulations. Type Ia supernovae\footnote{A Type Ia supernova is spectroscopically
defined by the absence of hydrogen emission lines, and is believed to occur in binary systems,
of which one of the stars is a white dwarf gradually accreting mass from its companion until a thermonuclear explosion takes place.} 
could also contribute to the production of 
$p$-nuclei, but its contribution remains affected by uncertainties in galactic chemical evolution models as well as in the
determination of the $s$-process enrichment prior to the $p$-process nucleosynthesis \cite{arnould2003,travaglio2011}. 

In addition to the uncertainties in the astrophysical modelling, significant uncertainties in the nuclear reaction rates 
bring further complications to our understanding of the nucleosynthesis. 
In order to estimate theoretically the abundance distribution, very large reaction-network calculations are required, 
typically including $\sim 2000$ nuclei and $\sim 20 000$ cross sections for the $p$-process, and 
$\sim 5000$ nuclei and $\sim 50 000$ cross sections for the $r$-process. Obviously, most of these cross sections
have not been determined experimentally and will not be measured in the near future. Hence, the calculations rely on
theoretical estimates of cross sections and reaction rates. These are usually derived within the 
Hauser-Feshbach theory~\cite{hauser},
where the main input parameters include optical-model potentials, nuclear level densities, and
$\gamma$-ray strength functions~(see e.g. Ref.~\cite{arnould2007,arnould2003} and references therein). 

\begin{figure*}[!t]
\begin{center}
\includegraphics[clip,width=1.7\columnwidth]{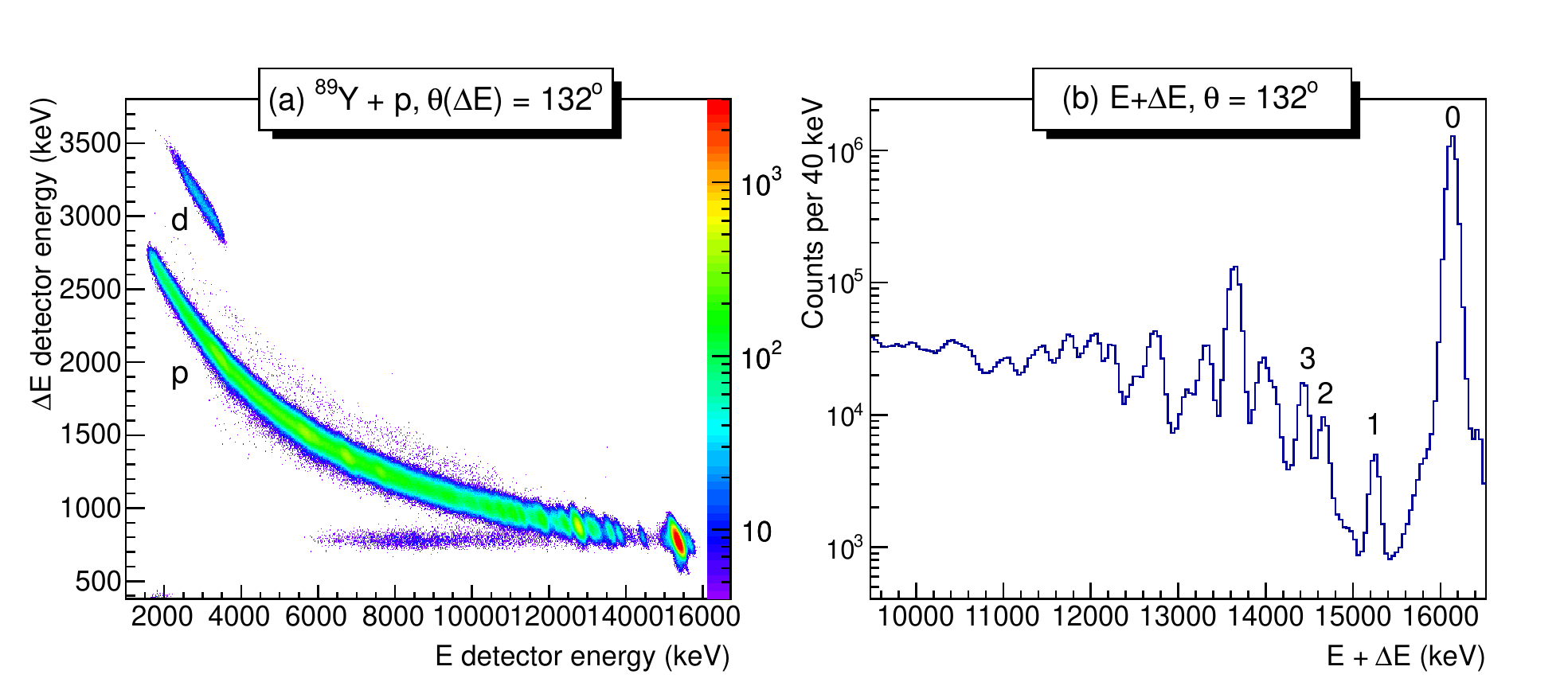}
\caption{(Color online) $\Delta E - E$ plot for $^{89}$Y$ + p$ (a), 
with the deposited energy in the thick $E$ detector versus the thin $\Delta E$ detector 
for $\theta = 132\pm1^{\circ}$, and (b) the sum of the deposited energy in the thin and thick detector for
the same angle. The labeled peaks are the ground state (0), the first excited 909-keV level (1), the second excited 
1507-keV level (2), and the third excited 1744-keV level (3).}
\label{fig:banana_matrices}
\end{center}
\end{figure*}

In this work, we have investigated the yttrium isotope $^{89}$Y by means of the reaction 
$^{89}$Y($p,p' \gamma$)$^{89}$Y, utilizing the Oslo method~\cite{schiller2000}
to extract the nuclear level density (NLD) and $\gamma$-ray strength function ($\gamma$SF). The NLD data
were previously reported in Ref.~\cite{guttormsen_yttrium_2014}, and this work focuses on the $\gamma$SF, and in particular the
enhancement at low-energy $\gamma$ rays. Extensive shell-model calculations are performed for the $M1$ strength, 
clearly showing a strong
increase at low $\gamma$-ray energies, in accordance with previous experimental~\cite{guttormsen2005,wiedeking2012} and 
theoretical~\cite{schwengner2013} findings in this mass region. We also investigate the impact of our results 
on capture cross sections and
astrophysical reaction rates. More specifically, we consider the cases $^{88}$Sr($p,\gamma$)$^{89}$Y and $^{88}$Y($n,\gamma$)$^{89}$Y, 
as these are relevant for the $p$-process in this mass region, and are also 
of interest for reaction networks in the context of stockpile stewardship~\cite{hoffman2006}.

The paper is organized as follows. In Sec.~\ref{sec:exp}, experimental details and an overview of the 
data analysis are given. 
Shell-model calculations and models for the $\gamma$-ray strength function are presented in Sec.~\ref{sec:shell}. 
In Sec.~\ref{sec:talys}, calculated cross sections and reaction rates are shown and compared with existing data. 
Finally, a summary and outlook can be found in Sec.~\ref{sec:sum}.

\section{Experimental details and extraction of level density and $\gamma$-strength function}
\label{sec:exp}

\subsection{Experimental details and unfolding of $\gamma$ spectra}
\label{sec:expdetails}
The experiment was performed at the Oslo Cyclotron Laboratory (OCL), 
utilizing a proton beam of 17 MeV.
The beam was impinging on a natural $^{89}$Y target with thickness 2.25 mg/cm$^2$.
The beam current was typically $\approx 0.5$ nA, with about 5 days of beam time,
including calibration runs on a natural Si target.

Charged ejectiles were measured with the 
Silicon Ring (SiRi) array~\cite{siri}, which is a $\Delta E-E$ telescope system
of 8 individual telescopes, each consisting of a 130-$\mu$m 8-fold segmented front
detector ($\Delta E$) with a $1500$-$\mu$m back detector ($E$) detector; hence, there are 64
individual silicon telescopes in the system. To reduce the amount of $\delta$ electrons from the target,
a 10.5-$\mu$m thick aluminium foil was placed in front of SiRi.
The SiRi system was mounted in
backward angles, covering the range $\theta = 126-140^{\circ}$ with an angular resolution of 
$\Delta \theta = 2^{\circ}$. Typical particle spectra from the experiment is shown in 
Fig.~\ref{fig:banana_matrices}. It is seen that the charged-particle species are 
clearly separated from each other. In the following, we gate on the protons, 
i.e. we consider the $^{89}$Y($p,p'\gamma$)$^{89}$Y reaction channel.
The proton-energy resolution was $\approx 130-160$ keV (FWHM), determined from the ground-state peak
and discrete peaks in the proton $\Delta E + E$ spectrum. 
 \begin{figure*}[!htb]
 \begin{center}
 \includegraphics[clip,width=2.1\columnwidth]{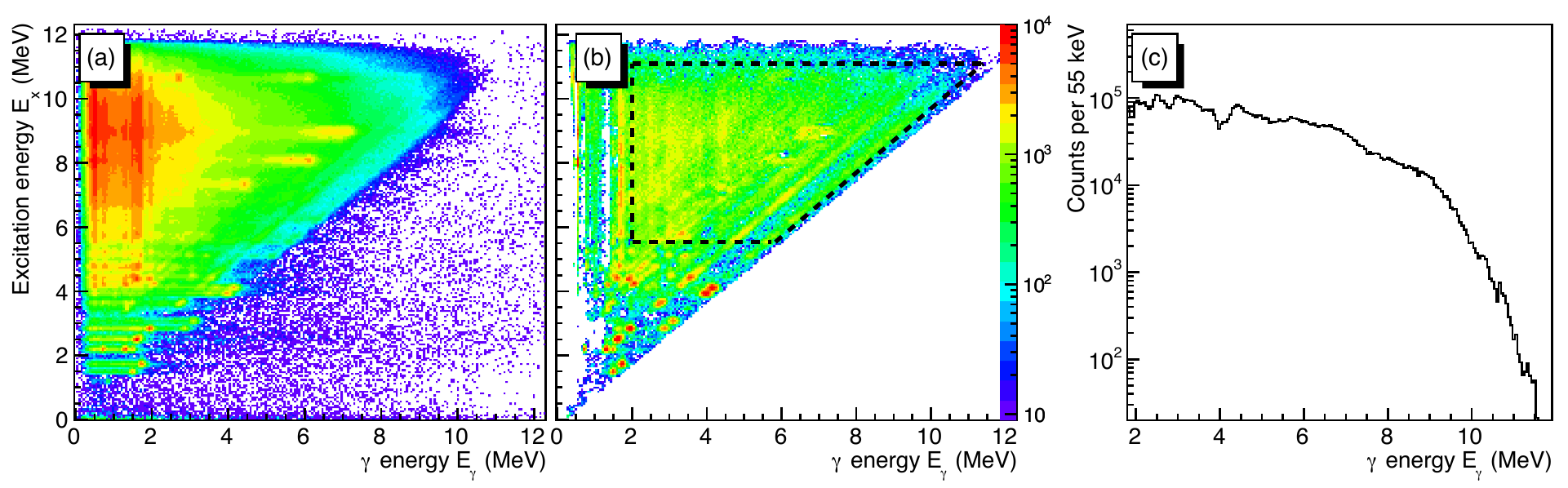}
\caption{(Color online) (a) Raw NaI spectra and 
(b) unfolded, primary $\gamma$ spectra for each initial excitation-energy bin for 
 $^{89}$Y($p, p' \gamma$)$^{89}$Y; the area within the dashed lines are used in the further analysis, i.e. 
 the data for  
 $E_{\gamma}>2.00$ MeV, $5.43 < E_x < 10.98$ MeV are selected.
 (c) Projection of the primary $\gamma$ spectra for the range of excitation energies between the lines.}
 \label{fig:fg_matrices}
 \end{center}
 \end{figure*}

The $\gamma$ rays were detected with the high-efficiency NaI(Tl) array CACTUS~\cite{CACTUS}. For this experiment,
CACTUS comprised 26 cylindrical NaI detectors of diameter 12.7 cm and length 12.7 cm 
mounted on the spherical CACTUS frame at angles 
$\Theta = 37.4^\circ, 63.4^\circ, 79.3^\circ, 100.7^\circ, 116.6^\circ$, and 142.6$^\circ$ with respect to the beam direction. 
All NaI crystals are collimated with lead cones to reduce the Compton contribution and enhance the peak-to-total ratio. 
The total efficiency for $E_\gamma = 1.33$ MeV was $\approx 14.1$\%. The trigger for the analog-to-digital converters 
of the CACTUS array was the signal from the thick $E$ detector in SiRi; this was also the start signal for the 
time-to-digital converters of the NaI detectors.

By selecting the proton channel and requiring the coincident $\gamma$-rays with the protons within a time 
window of $\approx 20$ ns, we obtain $\gamma$ spectra for each excitation energy, which is found from the 
measured proton energy in SiRi corrected for the reaction kinematics. The excitation-energy versus 
NaI signals are shown in Fig.~\ref{fig:fg_matrices}a. Some pileup events are observed for $E_{\gamma} > E_x$,
however, they are rare and do not contribute significantly. We also see contaminant $\gamma$ lines
from $^{12}$C and $^{16}$O, e.g., at $E_{\gamma} = $ 4.4, 6.1 and 7.1 MeV. These lines are kept in the 
data set for unfolding, and later removed as explained in the following.

The $\gamma$-ray
spectra for each excitation energy bin were corrected for the CACTUS response functions, 
i.e. removing the Compton, single-escape, double escape,
and back-scattered annihilation events, and correcting for the full-energy efficiency. The contamination lines
were also unfolded to obtain only the full-energy peaks in the final, unfolded  $\gamma$-ray
spectra. Then, a 3D scan of the area around the full-energy peaks was performed to 
get a correct estimate of the underlying spectrum, and finally removing the carbon and oxygen lines.
Experimental response functions have been recorded in-beam for $\gamma$ transitions of
$^{13}$C, $^{16}$O, $^{28}$Si, and $^{56}$Fe. The unfolding procedure is described in detail in Ref.~\cite{guttormsen1996};
the main advantage of this method is that the original, statistical fluctuations are preserved without introducing 
spurious fluctuations. This is obtained by applying a strong smoothing on the Compton background spectrum before
subtracting it from the raw spectrum. This approach is justified by the fact that the Compton background 
varies slowly with $\gamma$-ray energy. Hence, we avoid spurious structures in the final, unfolded spectrum.

The experimental spin range for the present experiment can be inferred from both the singles-proton spectrum 
and the proton-$\gamma$ coincidences. 
From the unfolded proton-$\gamma$ coincidences, we clearly identify transitions 
for states with spin/parity up to 7/2$^+$.
From the singles-proton spectrum, we see that the 909-keV isomer with spin/parity 
9/2$^+$ is populated. This provides a lower limit for the populated spins. 

\subsection{Primary $\gamma$-ray spectra and functional form of the level density and $\gamma$-ray strength}
\label{sec:rhosigchi}

Once the $\gamma$-spectra were properly unfolded, an iterative method~\cite{guttormsen1987} was applied to obtain the distribution
of primary $\gamma$ transitions from secondary and higher-order transitions. The principle of this method is that for a given 
excitation-energy bin $j$, this will contain all the $\gamma$-rays of the decay cascades from lower-lying bins $i<j$, and in addition the 
primary transitions for bin $j$. Thus, by subtracting a weighted sum of the spectra below $j$, the distribution of primary $\gamma$ rays 
for bin $j$ is obtained. Systematic uncertainties of this procedure is discussed in detail in Ref.~\cite{larsen2011}. The obtained
matrix of primary $\gamma$-ray spectra is displayed in Fig.~\ref{fig:fg_matrices}b, and the projection of the primary $\gamma$ rays
for $E_x\approx 5.5-11.0$ MeV is shown in Fig.~\ref{fig:fg_matrices}(b). We note that the spectrum in Fig.~\ref{fig:fg_matrices}(b)
is rather smooth, although some structures appear particularly at lower $E_\gamma$ values.  

The data within the dashed lines are used for the 
extraction of NLD and $\gamma$SF in the next step. The boundaries are chosen for the following reasons: we need to make sure 
that the decay originates from a region of fairly high level density, 
and also that the primary $\gamma$ spectra are indeed correct. An indicator for how well the primary-$\gamma$ extraction
procedure works, is calculated for each iteration and for each excitation-energy bin. In short, this indicator shows whether
the primary spectrum corresponds to a $\gamma$ multiplicity of 1, which it obviously should, and in such cases the indicator
is unity. We allow for a variation in this indicator of $\pm 15$\%; if the deviation from unity is larger, we do not
use the primary spectra from that excitation-energy region (see Ref.~\cite{guttormsen1987} for more details).
Therefore, cuts are made in the matrices as shown in Fig.~\ref{fig:fg_matrices}, ensuring relatively high initial excitation 
energies and thus high initial level density: $E_x = 5.44-10.97$ MeV, and $E_\gamma^{\mathrm{low}} = 2.01$ MeV for $^{89}$Y.
The $E_\gamma^{\mathrm{low}}$ limit is necessary because of strong, discrete transitions being either subtracted too little or 
too much below this energy, resulting in vertical "ridges" or "valleys" in the primary $\gamma$ matrix (see also the discussion
in Ref.~\cite{larsen2011}). 

To extract the NLD and the $\gamma$SF from the set of primary $\gamma$-ray spectra, we make use of the following 
relation~\cite{schiller2000}:
\begin{equation}
P(E_{\gamma},E_x) \propto  \rho (E_{\mathrm{f}}) {\mathcal{T}}  (E_{\gamma}).
\label{eq:brink}
\end{equation}
Here, $P(E_{\gamma},E_x)$ is the experimental primary $\gamma$-ray matrix as shown in Fig.~\ref{fig:fg_matrices}, but where 
the primary $\gamma$ spectra of each excitation-energy bin are normalized to unity to represent the probability of decay
from that bin~\cite{schiller2000}. 
The matrix of primary $\gamma$ spectra, $P(E_{\gamma},E_x)$, is proportional to the level density at the final 
excitation energy $E_f = E_x - E_{\gamma}$, and to the $\gamma$-ray transmission coefficient ${\mathcal{T}}  (E_{\gamma})$. 
The latter is dependent on the $\gamma$-ray energy only, in accordance with the generalized form of the Brink-Axel 
hypothesis~\cite{brink1955,axel1962}. The generalized Brink-Axel hypothesis has very recently been experimentally 
verified for $\gamma$ transitions in the quasicontinuum~\cite{guttormsen2016}.
The expression in Eq.~(\ref{eq:brink}) is valid for statistical decay, i.e. where the decay is independent
of the formation of the compound state~\cite{bohr-mottelson1969}.

The functional form of the NLD and $\gamma$SF is determined through a least-$\chi^2$ fit to the $P(E_{\gamma},E_x)$
matrix as described in Ref.~\cite{schiller2000}. The absolute normalization of the functions remains to be found, i.e. 
determining the parameters $\mathcal{A}$, $\mathcal{B}$, and $\alpha$ in the transformations
\begin{eqnarray}
\rho(E_x-E_\gamma)&=&\mathcal{A}\exp[\alpha(E_x-E_\gamma)]\,\tilde{\rho}(E_x-E_\gamma),
\label{eq:array1}\\
{\mathcal{T}}(E_\gamma)&=&\mathcal{B}\exp(\alpha E_\gamma)\tilde{{\mathcal{T}}} (E_\gamma),
\label{eq:array2}
\end{eqnarray}
which all give equally good fits to the experimental data. This normalization will be described in the following. 

\subsection{Normalization of the level densities}
\label{sec:nld}

\begin{table*}[ht]
\begin{center}
\caption{Neutron resonance parameters $D_0$ and $\left< \Gamma_{\gamma 0}\right>$ from Ref.~\cite{RIPL3}, and spin cutoff parameters from global systematics of
    Refs.~\cite{egidy2006,egidy2009};
    $A_f$ is the final nucleus following neutron capture, $J_t$ is the ground-state spin of the target nucleus, $S_n$ is the neutron-separation energy, 
    $\sigma_{05,09}$ are the spin-cutoff parameters from Eqs.~(\ref{eq:spincut}) and (\ref{eq:spincut05}), $D_0$ is the $s$-wave level spacing~\cite{RIPL3},
    and $\rho_{05,09}(S_n)$, are the
    total level densities calculated from $\sigma_{05,09}$. Finally, $\rho^{\mathrm{syst}}_{05,09}$ are the
    total level densities at $S_n$ as predicted from the global systematics of Refs.~\cite{egidy2006,egidy2009}, respectively. }
\begin{tabular}{lcclcllclll}
\hline
\hline
A$_f$      & $J_t$     & $S_n$ & $D_0$      & $\sigma_{05}(S_n)$ & $\rho_{05}(S_n)$      & $\rho^{\mathrm{syst}}_{05}(S_n)$ & $\sigma_{09}(S_n)$ & $\rho_{09}(S_n)$      & $\rho^{\mathrm{syst}}_{09}(S_n)$ & $\left< \Gamma_{\gamma 0}\right>$   \\
		   &           & (MeV) &  (keV)     &                    & (10$^{4}$ MeV$^{-1}$) & (10$^{4}$ MeV$^{-1}$)            &                    & (10$^{4}$ MeV$^{-1}$) & (10$^{4}$ MeV$^{-1}$)            & (meV)                               \\
\hline
$^{86}$Rb  & 5/2       & 8.651 & 0.172(8)   & 5.12               & 6.11(28)            &   21.6                          & 4.02               & 4.22(20)            & 12.4                            & 250(10)                 \\
$^{88}$Rb  & 3/2       & 6.083 & 1.630(150) & 4.80               & 0.78(7)             &   1.47                          & 3.85               & 0.53(5)             & 1.14                            & 170(30)    \\
$^{85}$Sr  & 0         & 8.530 & 0.320(120) & 4.89               & 15.3(57)            &   20.4                          & 3.91               & 9.88(370)           & 7.88                            & 240(80)    \\
$^{87}$Sr  & 0         & 8.424 & 2.600(800) & 5.07               & 2.02(62)            &   13.2                          & 3.94               & 1.23(38)            & 4.01                            & 260(80)    \\
$^{88}$Sr  & 9/2       & 11.11 & 0.290(80)  & 5.22               & 3.00(83)            &   13.1                          & 4.02               & 2.44(67)            & 6.40                            & 150(40)    \\
$^{89}$Sr  & 0         & 6.359 & 23.70(290) & 4.72               & 0.19(2)             &   0.53                          & 3.76               & 0.12(2)             & 0.39                            & 190(50)    \\
$^{90}$Y   & 1/2       & 6.857 & 3.700(400) & 4.99               & 0.70(8)             &   2.25                          & 3.93               & 0.44(5)             & 1.71                            & 130(40)    \\
$^{91}$Zr  & 0         & 7.194 & 6.000(1400)& 4.95               & 0.83(19)            &   1.20                          & 3.88               & 0.52(12)            & 0.91                            & 130(20)    \\
$^{92}$Zr  & 5/2       & 8.635 & 0.550(100) & 5.03               & 1.85(34)            &   4.11                          & 3.95               & 1.29(23)            & 2.60                            & 140(40)    \\
$^{93}$Zr  & 0         & 6.734 & 3.500(800) & 4.84               & 1.37(31)            &   1.69                          & 3.86               & 0.88(20)            & 1.31                            & 135(25)    \\
$^{94}$Zr  & 5/2       & 8.221 & 0.302(75)  & 4.95               & 3.29(82)            &   5.62                          & 3.94               & 2.34(58)            & 3.63                            & 157(20)    \\
$^{95}$Zr  & 0         & 6.462 & 4.000(800) & 4.79               & 1.17(24)            &   1.99                          & 3.86               & 0.77(15)            & 1.54                            & 85(20)    \\
$^{97}$Zr  & 0         & 5.575 & 13.00(300) & 4.66               & 0.34(8)             &   0.58                          & 3.76               & 0.23(5)             & 0.72                            & 65(15)    \\
\hline
\hline
\end{tabular}
\label{tab:par1}
\end{center}
\end{table*}

To determine the absolute scaling $\mathcal{A}$ and the slope $\alpha$ of the level density, our data points are normalized 
to discrete levels~\cite{NNDC} at low excitation energy. At the neutron separation energy $S_n$, data on average $s$-wave
neutron-resonance spacings $D_0$~\cite{RIPL3}, are usually used to calculate the total level density $\rho(S_n)$ 
for all spins and both parities.
For the case of $^{89}$Y, there are no neutron-resonance data available as $^{88}$Y is unstable.
To estimate reasonable normalization parameters we have considered systematics of $s$-wave ($\ell=0$) 
resonance spacings for this mass region
using the most recent evaluation of the Reference Input Parameter Library (RIPL-3, Ref.~\cite{RIPL3}). 
In addition, systematic errors due to the spin distribution at $S_n$ must be 
taken into account~\cite{larsen2011}. This will be discussed in the following.

\begin{figure}[t]
 \begin{center}
 \includegraphics[clip,width=1.\columnwidth]{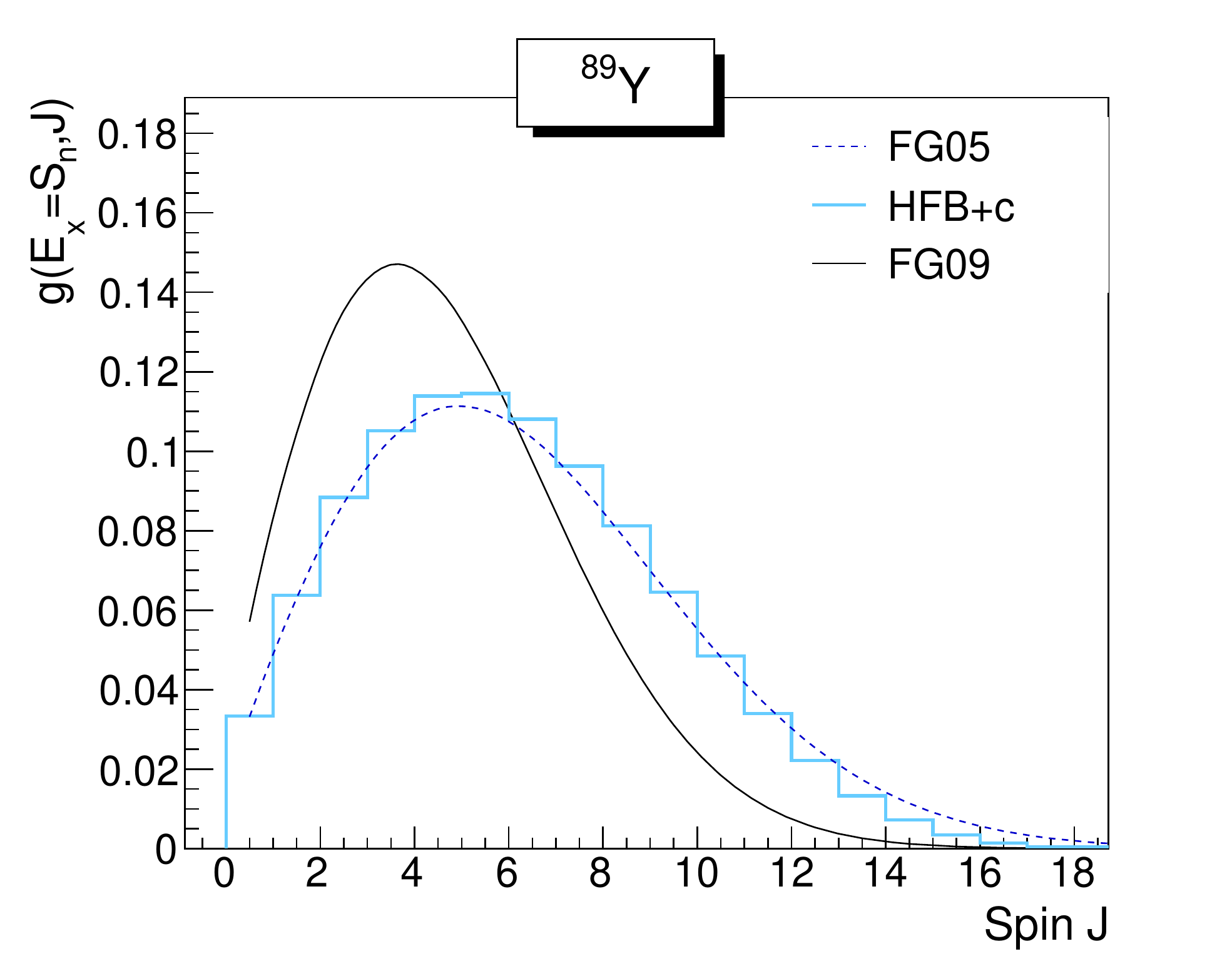}
 \caption{(Color online) Spin distributions for $^{89}$Y at the neutron separation energy $S_n = 11.482$ MeV
for the three different normalization approaches. For FG09 and FG05, Eq.~(\ref{eq:spindistr}) is used with their
respective spin cutoff parameters in Eqs.~(\ref{eq:spincut}) and (\ref{eq:spincut05}). The HFB+c calculations
assume no specific shape of the spin distribution, but happens to be very similar to FG05 in this case.}
 \label{fig:spins}
 \end{center}
 \end{figure}

In our recent work~\cite{guttormsen_yttrium_2014}, the $^{89,90}$Y level densities were normalized using global systematics
of Ref.~\cite{egidy2009} within the constant-temperature (CT) approximation of the level-density function. 
This approach gives a very good description of the functional form
of experimental level densities above $\approx 2\Delta$~\cite{moretto2015}, where $\Delta\approx 1$ MeV is the pair-gap parameter. 
However, it yields rather low (and constant) values for the spin-cutoff parameter. 
Moreover, this is only one out of several possible values for the spin-cutoff parameter and hence for the upper 
normalization point.
Also, as the spin distribution for excitation energies up to the neutron separation energy is needed for the normalization of the $\gamma$SF 
later on, we will here rely on an excitation-energy dependent spin-cutoff parameter. We have chosen three different approaches, the first
one using a phenomenological Fermi-gas (FG) spin cutoff parameter following Ref.~\cite{egidy2009} (FG09):
\begin{equation}
\sigma_{09}^2(E_x) = 0.391 A^{0.675}(E_x - 0.5Pa^{\prime})^{0.312}. 
\label{eq:spincut}
\end{equation}
Here, $E_x$ is the excitation energy, $A$ is the mass number and $Pa^{\prime}$ is the deuteron pairing energy 
as defined in Ref.~\cite{egidy2009}.
Secondly, we consider the
FG spin cutoff parameter of Ref.~\cite{egidy2006}, where a rigid-body moment of inertia is assumed (FG05):
\begin{equation}
\sigma_{05}^2(E_x) = 0.0146A^{5/3}\frac{1+\sqrt{1+4a(E_x-E_1)}}{2a}.
\label{eq:spincut05}
\end{equation}
Here, $a$ is the level-density parameter and $E_1$ is the excitation-energy backshift determined from global systematics of
Ref.~\cite{egidy2006}. 
For the phenomenological spin cutoff parameters, the spin distribution is given by the standard expression~\cite{ericson1959,ericson1960}:
\begin{equation}
g(E_x,J) \simeq \frac{2J+1}{2\sigma^2(E_x)} \exp\left[ \frac{-(J+1/2)^2}{2\sigma^2(E_x)}\right].
\label{eq:spindistr}
\end{equation}
Using the phenomenological spin cutoff parameters and assuming equiparity
(as shown in Ref.~\cite{guttormsen_yttrium_2014}),
the total level density at $S_n$ can be estimated from $D_0$
through the expression~\cite{schiller2000}
\begin{equation}
\rho(S_n) = \frac{2\sigma^2}{D_0} \cdot \frac{1}{(J_t+1)\exp\left[-(J_t+1)^2/2\sigma^2\right] + J_t\exp\left[-J_t^2/2\sigma^2\right]},
\label{eq:rhoSn}
\end{equation}
where $J_t$ is the ground-state spin of the target nucleus in the neutron-resonance experiment.
Finally, we use microscopic calculations within the Hartree-Fock-Bogoliubov plus combinatorial (HFB+c) approach~\cite{goriely2008}.
The three spin distributions are shown in Fig.~\ref{fig:spins} for the case of $^{89}$Y.
 \begin{figure*}[!hbt]
 \begin{center}
 \includegraphics[clip,width=1.8\columnwidth]{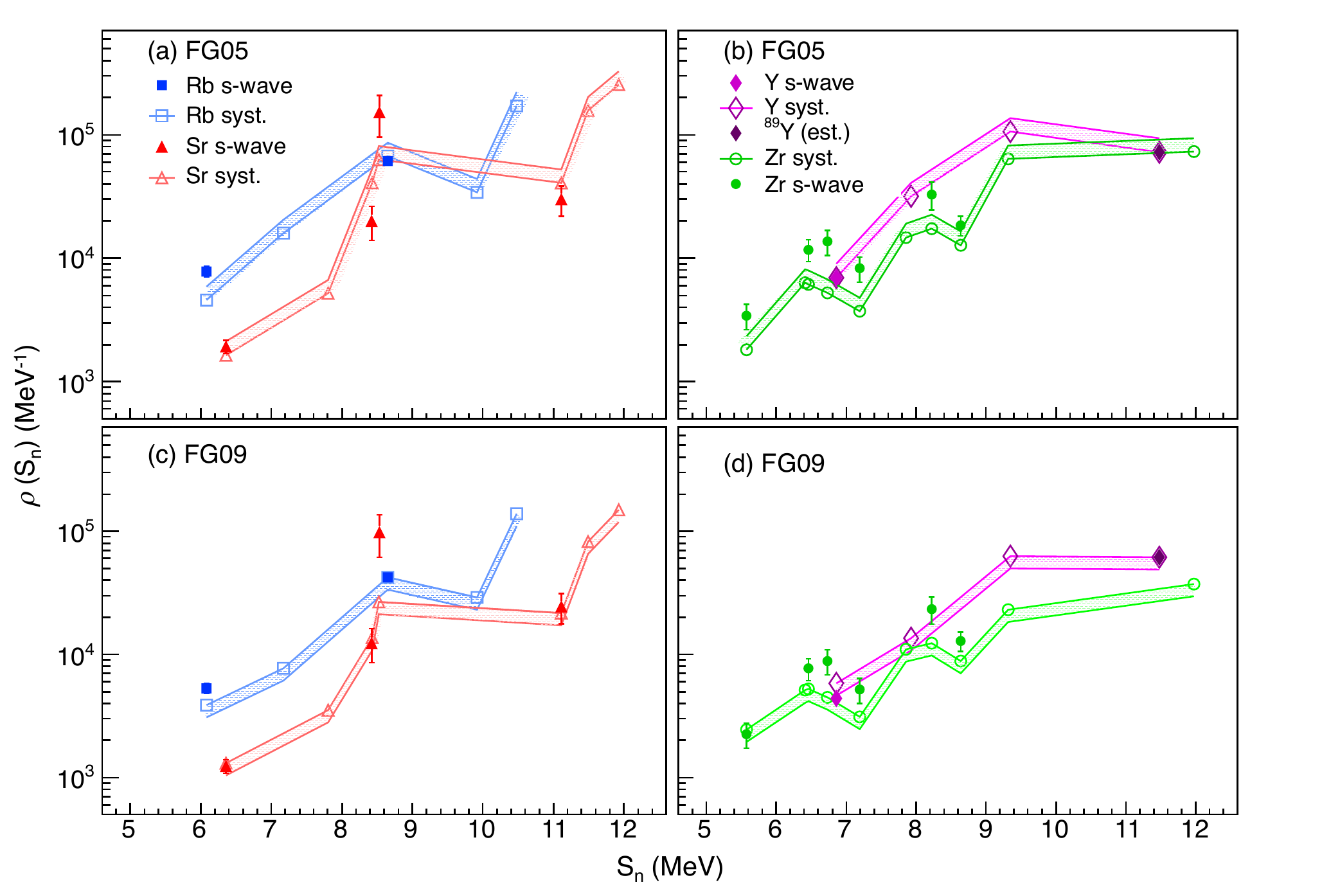}
 \caption{(Color online) Level densities at $S_n$ in the Y mass region
 with global systematics of (a-b) Ref.~\cite{egidy2006} (FG05)
 and (c)-(d) Ref.~\cite{egidy2009} (FG09)
 (see Table~\ref{tab:par1}). The unknown level density for $^{89}$Y is shown as a purple diamond.
 The global-systematics predictions are scaled with a factor of 0.31 and 0.34 for the FG05 and FG09 approaches,
 respectively. The error bands show the upper-limit scaling of 0.40 for FG05 (a-b) and the lower limit of 0.27 for FG09 (c-d). }
 \label{fig:rhosyst_FG}
 \end{center}
 \end{figure*}

We observe that the spin distribution of Eq.~(\ref{eq:spindistr}) using $\sigma_{05}$ is very broad and centered at significantly
higher spins than using $\sigma_{09}$. We consider therefore $\sigma_{05}$ to give the upper limit, and $\sigma_{09}$
as the lower limit in estimating $\rho(S_n)$ for $^{89,90}$Y. The calculated $\rho(S_n)$ values for Rb, Sr, Y, and Zr isotopes 
are given in Table~\ref{tab:par1} together with the applied input parameters.
The resulting systematics for the level densities at $S_n$ are shown in Fig.~\ref{fig:rhosyst_FG}.
Note that the predictions from the global systematics
are fitted to the semi-experimental data points through a common scaling factor of
$0.31^{+0.09}_{-0.10}$ and $0.34^{+0.08}_{-0.07}$
for $\rho_{05,09}(S_n)$, respectively. Using the upper (lower) $\chi^2$ uncertainty
for the $\sigma_{05}$ ($\sigma_{09})$ results,
we get the following
estimates for $^{89}$Y ($S_n=11.482$ MeV):
$\rho^{\mathrm{low}}_{09}(S_n) = 4.87 \cdot 10^4$~MeV$^{-1}$,
corresponding to $D_0 = 143$ eV for $\sigma_{09}(S_n)$ = 4.12, and $\rho^{\mathrm{up}}_{05}(S_n) = 9.33 \cdot 10^4$ MeV$^{-1}$,
corresponding to $D_0 = 100$ eV for $\sigma_{05}(S_n)$ = 5.45. Finally, the HFB+c calculations (with no
excitation-energy shift, $\delta = 0.0$ MeV where $\delta$ is defined in Ref.~\cite{goriely2008}) 
yield $\rho_{\mathrm{HFB+c}}(S_n) = 7.09 \cdot 10^4$~MeV$^{-1}$
with $D_0 = 121$ eV. 
A fit of Eq.~(\ref{eq:spindistr}) on the HFB+c calculations at $S_n$ gives $\sigma_{\mathrm{HFB+c}} = 4.89$,
i.e. in between the two phenomenological approaches.

Due to the cutoff on the $\gamma$ energy, $E_\gamma^{\mathrm{low}}$(see Fig.~\ref{fig:fg_matrices}), 
our level-density data points will reach a maximum excitation energy of
$S_n - E_\gamma^{\mathrm{low}}$. Following Ref.~\cite{guttormsen_yttrium_2014}, we use an interpolation between our data points 
up to $\rho(S_n)$ of the constant-temperature form~\cite{ericson1959}:
\begin{equation}
\rho_{CT}(E_x) = \frac{1}{T}\exp{\frac{E_x-E_0}{T}},
\label{eq:rhoCT}
\end{equation}
where $E_0$ is the excitation-energy shift and $T$ is the constant nuclear temperature. 
For $^{89}$Y, we have used $(E_0,T)$ = (0.658,0.95),
(0.069,1.02), and (0.098,1.05) MeV for FG05, HFB+c, and FG09 respectively.
 \begin{figure}[t]
 \begin{center}
 \includegraphics[clip,width=1\columnwidth]{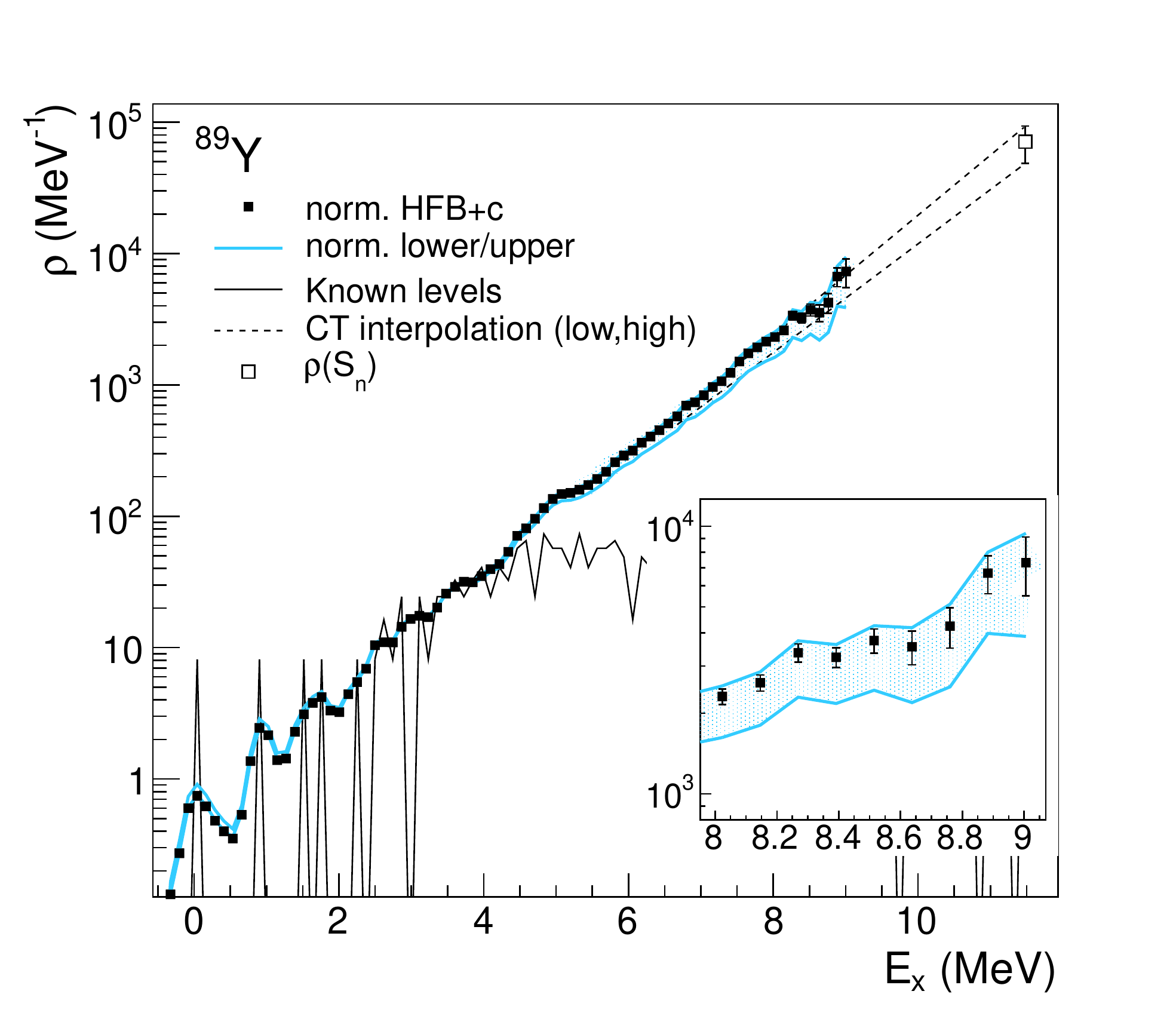}
 \caption{(Color online) The extracted level density of $^{89}$Y. The black points give the  
 HFB+c normalization, the lower and upper limits are shown as azure lines.
 The insert shows a zoom of the high-$E_x$ part.}
 \label{fig:nld}
 \end{center}
 \end{figure}

The normalized level density is shown in Fig.~\ref{fig:nld}. Our data points follow the discrete levels well up to
$E_x \approx 4.5$ MeV, which means that these levels are populated by primary $\gamma$ decay from the higher-lying levels
in the region; above this energy the level densities increase rapidly,
while the known levels show a saturation when reaching $\approx 65$ MeV$^{-1}$. As discussed in Ref.~\cite{guttormsen_yttrium_2014},
the level density displays a remarkable constant-temperature behavior. The lower and
upper limits for the normalization are also shown, representing the systematic errors. Note that the lower limit
is very similar to the normalization used in Ref.~\cite{guttormsen_yttrium_2014} (within 18\%). 
At low excitation energies there is obviously not much difference between the normalization options, 
as our data points are fixed to the discrete levels. 
At the neutron separation energy the systematic uncertainty is at its maximum, within a factor of $\approx 2$ 
(see insert of Fig.~\ref{fig:nld}).

\subsection{$\gamma$-ray strength function}
\label{sec:gsf}

The slope of the $\gamma$SFs is determined through the normalization of the level densities (see Eqs.~(\ref{eq:array1}) and (\ref{eq:array2})). 
The absolute scale $\mathcal{B}$ can be found by use of the total, average radiative width $\left< \Gamma_{\gamma0} \right>$. 
The average radiative width of neutron s-wave capture resonances with spins 
$J_t \pm 1/2$ expressed in terms of the experimental ${\mathcal T}$ is given by~\cite{voinov2001}:
\begin{align}
\langle \Gamma_{\gamma0}(S_n,J_t\pm 1/2,&\pi_t)\rangle =
 \frac{\mathcal{B}}{4\pi\rho(S_n,J_t\pm 1/2,\pi_t)}\int_{E_{\gamma}=0}^{S_n}\mathrm{d}E_{\gamma}\mathcal{T}(E_{\gamma}) \nonumber \\ 
 &\times \rho(S_n-E_{\gamma}) \sum_{J= -1}^{1} g(S_{n}-E_{\gamma},J_{t}\pm 1/2+J),
\label{eq:width}
\end{align}
where $J_t$ and $\pi_t$ are the spin and parity of the target nucleus in the $(n,\gamma)$ reaction, 
$\rho(S_n-E_{\gamma})$ is the normalized, experimental level density obtained in Sec.~\ref{sec:nld}, and
$\mathcal{T}$ is the experimental transmission coefficient, which in principle includes all types of electromagnetic transitions:
$\mathcal{T}_{E1}+\mathcal{T}_{M1}+\mathcal{T}_{E2}+ ...$ .
The sum runs over all final states with spins $J_t \pm 1/2 + J$, where $J= -1,0,1$ from considering the spins 
reached after one primary dipole transition with energy $E_\gamma$ (see also Eq.~(3.1.) in Ref.~\cite{kopecky_uhl_1990}).  
Note that the factor
$1/\rho(S_n,J_t\pm 1/2,\pi_t)$ equals the neutron resonance spacing $D_0$. From the normalized transmission coefficient,
the $\gamma$SF is determined by
\begin{equation}
f(E_\gamma) = \frac{\mathcal{T}(E_{\gamma})}{2\pi E_\gamma^3},
\end{equation}
using the fact that dipole transitions dominate the strength for the considered $E_x$ region~\cite{kopecky_uhl_1990,larsen2013}.

For $^{89}$Y, we have estimated the unknown $\left< \Gamma_{\gamma0} \right>$ from data in this mass region~\cite{RIPL3}.
Specifically, we took the average value of the nuclei close in mass to $^{89}$Y, namely
$^{88,89}$Sr, $^{90}$Y, and $^{91}$Zr, see Table~\ref{tab:par1}. With an uncertainty of $\approx$ 25\%, we obtain 
$\left< \Gamma_{\gamma0} \right>(^{89}\mathrm{Y}) = 150(38)$ meV.
The assumed uncertainty of 25\% is strongly guided by the comparison with photo-nuclear data from the reactions
$^{89}$Y($\gamma,n$)+$^{89}$Y($\gamma,np$)~\cite{berman1967,lepretre1971,varlamov2003}. 
These cross-section data are converted into $\gamma$SF by the relation~\cite{RIPL3}
\begin{equation}
f(E_\gamma) = \frac{\sigma_{(\gamma,n)}(E_\gamma)}{3\pi^2 \hbar^2 c^2 E_\gamma},
\end{equation}
again assuming that dipole radiation is dominant, which is reasonable in this $E_\gamma$ region (see e.g.~\cite{RIPL3}
and references therein).
 \begin{figure}[!t]
 \begin{center}
 \includegraphics[clip,width=1\columnwidth]{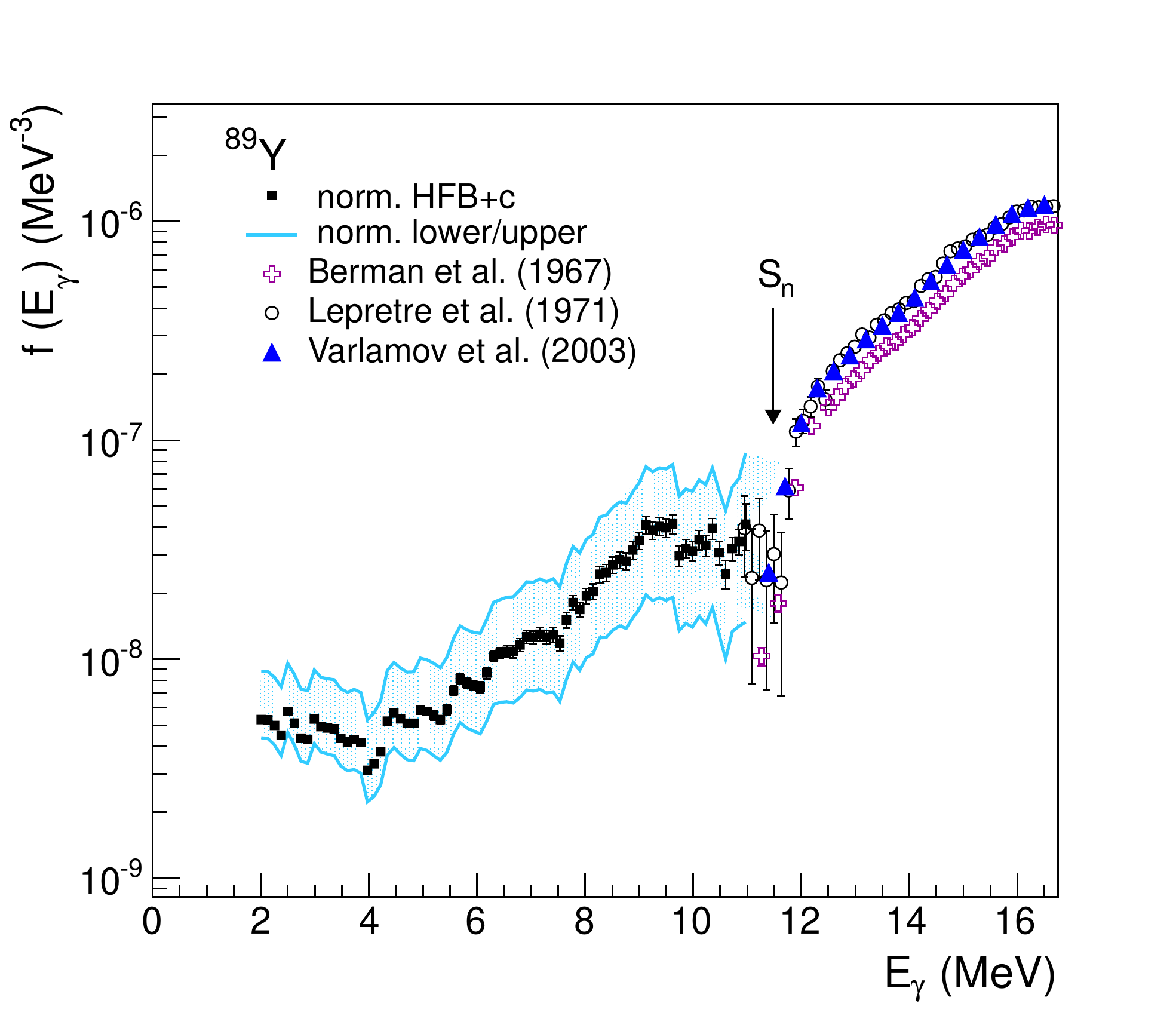}
 \caption{(Color online) Normalized $\gamma$SF of $^{89}$Y compared to $^{89}$Y($\gamma,n$)+$^{89}$Y($\gamma,np$) data from 
 Refs.~\cite{berman1967,lepretre1971}, and evaluated data from Ref.~\cite{varlamov2003}. 
 The black points are obtained with the HFB+c 
 normalization, the azure lines show the lower and upper limits.}
 \label{fig:gsfdata}
 \end{center}
 \end{figure}

The normalized $\gamma$SF is shown in Fig.~\ref{fig:gsfdata}. The error bands include the uncertainty in
$\left< \Gamma_{\gamma0} \right>$ as well as in the level density and the choice of spin distribution. The 
uncertainty in absolute value ranges from a factor of 1.8 at $E_\gamma = 2.0$ MeV to $\approx 4.8$ 
at $E_\gamma = 11.3$ MeV.

For $\gamma$-ray energies above $3-4$ MeV, we find that the
strength is increasing as a function of $\gamma$-ray energy, as expected for the tail of the giant dipole resonance
(GDR)~\cite{dietrich1988}. We also observe a drop in strength for $\gamma$ rays between 
$\approx 9.6-11.0$ MeV. This is understood by looking at the primary $\gamma$-ray matrix in Fig.~\ref{fig:fg_matrices}a,
where it is clear that the upper right corner in the triangle has significantly fewer statistics than for lower
$\gamma$-ray energies. This could be due to less direct feeding to the ground state, with spin/parity $1/2^-$, and
more to the first excited $9/2^+$ state at 909 keV. We note that this behavior is in agreement with the ($\gamma,\gamma^{\prime}$) 
data from Ref.~\cite{benouaret2009}, which also display a reduction in intensity in the energy range 
$E_\gamma \approx 9.8 - 11.3$ MeV. This indicates that in this particular region, the extracted data are not representative
for a general $\gamma$SF, because there is, very likely, a strong dependence on the final state(s) and the individual
overlap with the initial and final state(s) of the transition(s).

Our data show an increase at decreasing $\gamma$ energies for $E_\gamma < 4$ MeV. 
This phenomenon, hereafter called the \textit{upbend}, has been subject of great interest recently, 
and was first discovered in iron isotopes about a decade ago~\cite{voinov2004}. 
In Ref.~\cite{larsen2011}, simulations on $^{57}$Fe using the DICEBOX code~\cite{DICEBOX} suggested that an enhancement
in the $E1$ strength could be invoked, although it was not present in the input $E1$ strength for the simulation. It was stressed that 
the low-energy enhancement could not be due to artifacts in the unfolding or the method for extracting primary transitions, 
as the same feature was seen using the primary transitions directly from the DICEBOX simulations. 
Three main reasons for this behavior was pointed out: 
\begin{itemize}
\item[(\textit{i})]{There was no significant contribution from quadrupole (E2) transitions
in the simulations;} 
\item[(\textit{ii})]{Specific restrictions were applied on the initial spin population;}
\item[(\textit{iii})]{The level density was very low for high spins in this particular simulation.}
\end{itemize}
Hence, the simulations indicated an increase in the low-energy part of the $\gamma$SF
in cases where the level density is low and the reaction populates selectively high spins and 
a rather narrow spin range.

However, this hypothesis was disproved in Ref.~\cite{wiedeking2012}, where a different and 
virtually model-independent technique was applied on data from the $^{94}$Mo(d,p$\gamma\gamma$)$^{95}$Mo
reaction. In contrast to the $^{96}$Mo($^{3}$He,$\alpha$)$^{95}$Mo data from Ref.~\cite{guttormsen2005},
where the reaction favors high-$\ell$ transfer, the (d,p) reaction mainly populates low-$\ell$ states. 
Nevertheless, the same shape of the $\gamma$SF was found~\cite{wiedeking2012}. 
Furthermore, using the Oslo method, 
it was shown in Ref.~\cite{larsen2013} that the $^{56}$Fe$(p,p^\prime)$$^{56}$Fe and 
the $^{57}$Fe($^3$He,$\alpha$)$^{56}$Fe data sets yielded very similar $\gamma$SFs.  
Therefore, considering the available data as of today, other explanations must be sought for 
to explain the upbend feature.

The upbend has recently been shown to be dominantly of dipole nature 
in $^{56}$Fe~\cite{larsen2013}
and in $^{151,153}$Sm~\cite{simon2016}, 
but the electromagnetic character is not known at present. 
Theoretical attempts suggest that both $E1$ as well as $M1$ strength may
contribute. An enhancement of $E1$ type is predicted from a thermal
coupling of quasiparticles to the continuum of unbound states at relatively
high temperature~\cite{litvinova2013}, whereas an enhancement of $M1$ strength is found in
shell-model calculations~\cite{schwengner2013,brown2014}. 
Moreover, the presence of the upbend may enhance 
astrophysical ($n,\gamma$) rates of exotic neutron-rich nuclei by up to 2 orders of magnitude~\cite{larsen2010}.
In the following, we will present shell-model calculations as well as models for the $E1$
strength, and compare with our data.

\section{Shell-model calculations and models for the dipole strength}
\label{sec:shell}

The shell-model calculations were performed by means of the code RITSSCHIL
\cite{zwa85} using a model space composed of the
$(0f_{5/2}, 1p_{3/2}, 1p_{1/2}, 0g_{9/2})$ proton orbits and the
$(0g_{9/2}, 1d_{5/2}, 0g_{7/2})$ neutron orbits relative to a $^{68}$Ni
core. This configuration space was also applied in our earlier study of
$M1$ strength functions in $^{94,95,96}$Mo and $^{90}$Zr \cite{schwengner2013}. In the
present calculations, two protons were allowed to be lifted from the $fp$
shell to the $0g_{9/2}$ orbit and two neutrons from the $0g_{9/2}$ to the 
$1d_{5/2}$ orbit. This resulted in dimensions up to 29000. 
The additional inclusion of the $\nu(0g_{7/2})$ orbit has negligible influence
on the low-energy part of the strength function, but produces a few strong
$M1$ transitions around 7 MeV dominated by the 
$\nu(0g_{9/2})^{-1} \nu(0g_{7/2})$ spin-flip configuration
(cf. Ref.~\cite{schwengner2013}). 
As these transitions do not describe the spin-flip peak completely up to high
energy, we use a phenomenological description for the spin-flip resonance in
the present work.

The calculations included states with spins from $J$ = 1/2 to 21/2 for
$^{89}$Y. 
For each spin the lowest 40
states were calculated. Reduced transition probabilities $B(M1)$ were
calculated for all transitions from initial to final states with
energies $E_f < E_i$ and spins $J_f = J_i, J_i \pm 1$. For the minimum and
maximum $J_i$, the cases $J_f = J_i - 1$ and $J_f = J_i + 1$, respectively,
were excluded. This resulted in more than 32000 $M1$ transitions for each
parity $\pi = +$ and $\pi = -$, which were sorted into 100 keV bins according
to their transition energy $E_\gamma = E_i - E_f$. The average $B(M1)$ value
for one energy bin was obtained as the sum of all $B(M1)$ values divided by the
number of transitions within this bin.

\begin{figure}[tb]
\begin{center}
\includegraphics[clip,width=1\columnwidth]{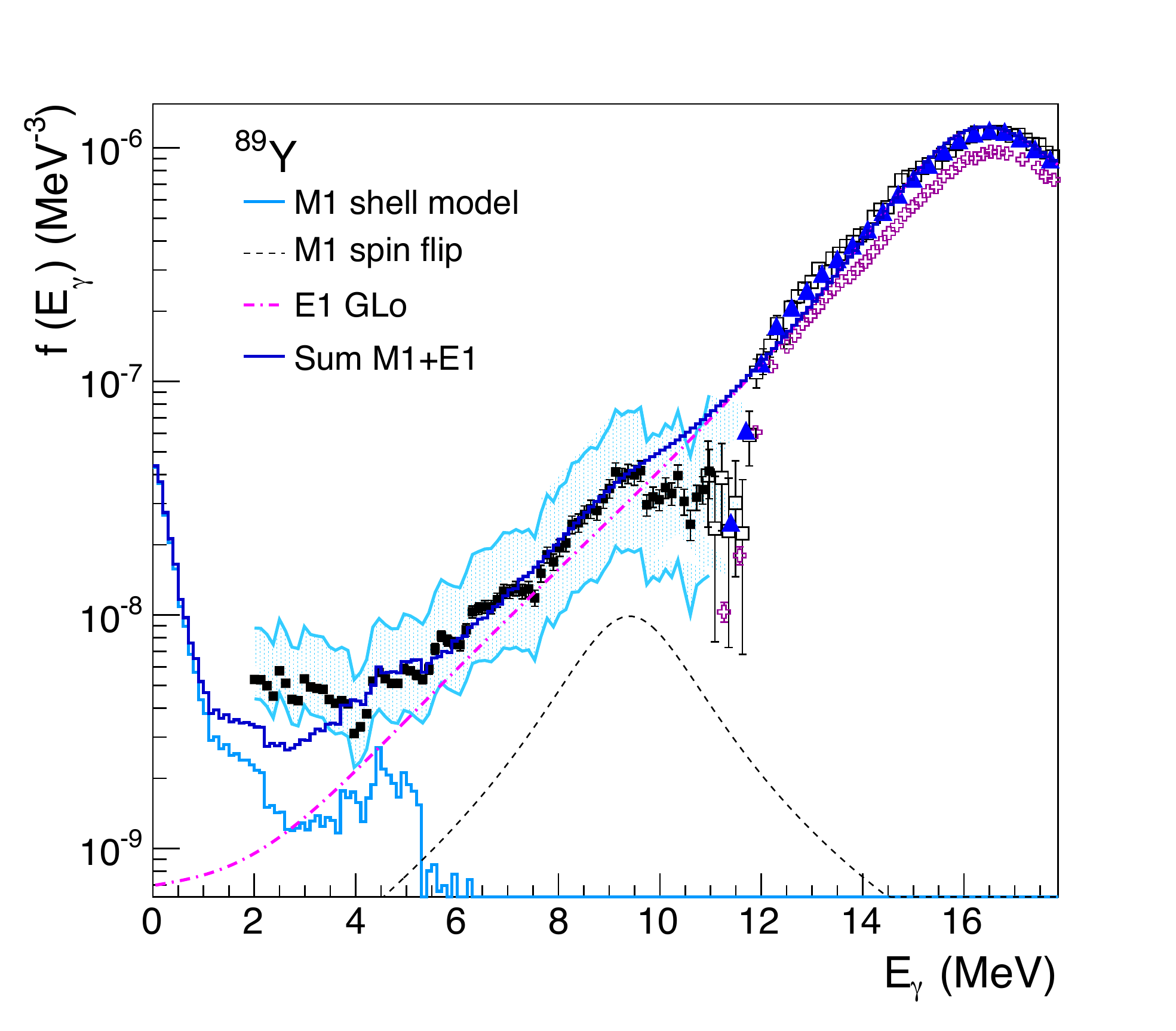}
\caption{(Color online) The $\gamma$SF of $^{89}$Y together with photo-nuclear data~\cite{berman1967,lepretre1971}
and evaluated ($\gamma,n$) data from Ref.~\cite{varlamov2003},
compared to models for the dipole strength.}
\label{fig:strength_calc}
\end{center}
\end{figure}

The $M1$ strength functions were deduced using the relation 
\begin{equation}
f_{M1}(E_\gamma) = \frac{16\pi}{9(\hbar c)^{3}} \overline{B}(M1,E_\gamma) \rho(E_i).
\end{equation}
They were calculated by multiplying the ${B(M1)}$ value in $\mu^2_N$ of each
transition with $11.5473 \times 10^{-9}$ times the level density at the energy
of the initial state $\rho(E_i)$ in MeV$^{-1}$ and deducing averages in energy
bins as done for the $\overline{B}(M1)$ values (see above). The level densities
$\rho(E_i,\pi)$ were determined by counting the calculated levels within energy
intervals of 1 MeV for the two parities separately. The strength functions
obtained for the two parities were subsequently added.
When calculating the strength functions, gates were set on the excitation
energy $E_x$ that correspond to the ones applied in the analysis of the
experimental data (see Sec.~\ref{sec:exp}).
The resulting $M1$ strength function for $^{89}$Y is shown
in Fig.~\ref{fig:strength_calc}. The low-energy
behavior of this strength function is very similar to that of the strength
functions calculated for the neighboring nuclei $^{94,95,96}$Mo, $^{90}$Zr
\cite{schwengner2013} and for $^{56,57}$Fe \cite{brown2014}.

The low-energy enhancement of $M1$ strength is caused by transitions between
many close-lying states of all considered spins located above the yrast line in
the transitional region to the quasi-continuum of nuclear states.  
Inspecting the wave functions, one finds large $B(M1)$ values for transitions
between states that contain a large component (up to about 50\%) of the same
configuration with broken pairs of both protons and neutrons in high-$j$
orbits, whereas states containing the unpaired $1p_{1/2}$ proton and proton
excitations only are not depopulated by strong $M1$ transitions. The largest
$M1$ matrix elements connect configurations with the spins of high-$j$ protons
re-coupled with respect to those of high-$j$ neutrons to the total spin
$J_f = J_i, J_i \pm 1$. The corresponding main configurations for
negative-parity states in $^{89}$Y are generated by exciting neutrons over the
shell gap at $N$ = 50, such as
$\pi(1p_{1/2}^1) \nu(0g_{9/2}^{-1} 1d_{5/2}^1)$ or 
$\pi(1p_{1/2}^1) \nu(0g_{9/2}^{-2} 1d_{5/2}^2)$ and by additional proton
excitations within the $(fp)$ shell, i.e.
$\pi[(0f_{5/2}, 1p_{3/2})^{-1} 1p_{1/2}^2] \nu(0g_{9/2}^{-1} 1d_{5/2}^1)$
and also proton excitations over the subshell gap at $Z$ = 40,
$\pi[(0f_{5/2}, 1p_{3/2})^{-1} 1p_{1/2}^0 0g_{9/2}^2]\nu(0g_{9/2}^{-1} 1d_{5/2}^1)$.
The positive-parity states require the excitation of an $(fp)$ proton to the
$0g_{9/2}$ orbit, for example 
$\pi(1p_{3/2}^{-1} 1p_{1/2}^1 0g_{9/2}^1) \nu(0g_{9/2}^{-1} 1d_{5/2}^1)$.
The orbits in these configurations have large $g$ factors with opposite signs
for protons and neutrons. Combined with specific relative phases of the proton
and neutron partitions they cause large total magnetic moments. 
\begin{figure}[tb]
\begin{center}
\includegraphics[clip,width=1\columnwidth]{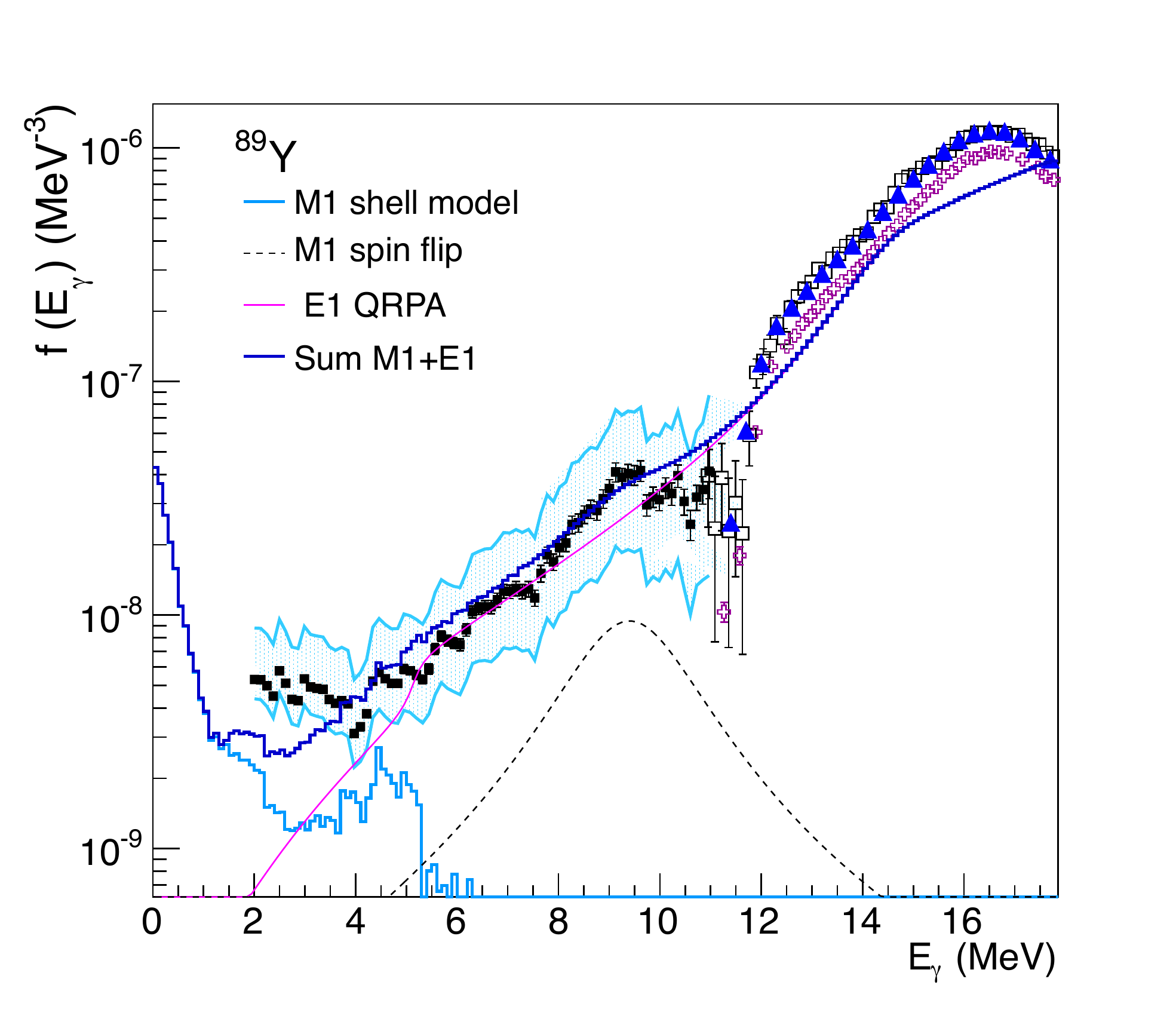}
\caption{(Color online) Same as Fig.~\ref{fig:strength_calc} but with the QRPA $E1$ strength (see text).}
\label{fig:strength_calc_QRPA}
\end{center}
\end{figure}

For a comparison with the experimental data, an $E1$ contribution to the
strength function has to be added. We have chosen two ways to estimate the $E1$ strength:
(\textit{i}) the phenomenological Generalized Lorentzian (GLo) model~\cite{kopecky_uhl_1990},
and (\textit{ii}) a microscopic approach based on the
quasiparticle-random-phase approximation (QRPA)~\cite{goriely2003,goriely2004}.
For option (\textit{i}), we apply the GLo
model with a constant temperature of the final states $T_f$, in contrast to a variable temperature
which depends on the final excitation energy. The choice of a constant temperature is in 
accordance with the Brink hypothesis~\cite{brink1955} and our ansatz that the $\gamma$-transmission
coefficient is, on average, independent of excitation energy in the statistical $E_x$ region. 
This is also in accordance with the constant-temperature level density found for $^{89}$Y. 
The GLo model is then given by
\begin{align}
& f_{\rm GLo}(E_{\gamma},T_f) = \frac{1}{3\pi^2\hbar^2c^2}\sigma_{E1}\Gamma_{E1} \times \\ \nonumber
& \left[\frac{ E_{\gamma} \Gamma(E_{\gamma},T_f)}{(E_\gamma^2-E_{E1}^2)^2 + E_{\gamma}^2 
	\Gamma (E_{\gamma},T_f)^2} + \;0.7\frac{\Gamma(E_{\gamma}=0,T_f)}{E_{E1}^3} \right],
\label{eq:GLO}
\end{align} 
with
\begin{equation}
\Gamma(E_{\gamma},T_f) = \frac{\Gamma_{E1}}{E_{E1}^2} (E_{\gamma}^2 + 4\pi^2 T_f^2).
\end{equation}
The parameters $\Gamma_{E1}$, $E_{E1}$ and $\sigma_{E1}$ correspond to the width, 
centroid energy, and peak cross section of the GDR.
For option (\textit{ii}), the $E1$ strength is obtained from large-scale QRPA calculations
on top of a Skyrme-Hartree-Fock-Bogoliubov description of the ground state. The 
QRPA calculations are performed in the spherical approximation, and a folding procedure is
applied to obtain the correct spreading width of the GDR. For more details on the QRPA calculations,
we refer the reader to Ref.~\cite{goriely2004}. The $E1$ calculations were taken from the
BRUSLIB library~\cite{BRUSLIB}.  

We included an $M1$ spin-flip resonance
with a standard-Lorentzian form~\cite{RIPL3}, using parameters in accordance with a recent
($p,p^\prime$)$^{90}$Zr experiment~\cite{iwamoto2012}. Strong $M1$ transitions were also observed in a photon-scattering experiment
in the excitation-energy region $\approx8-10$ MeV~\cite{rus13}.
For the QRPA calculation, we had to shift the $E1$ strength
with an energy shift of $\delta_{\mathrm{QRPA}} =+1.5$ MeV, i.e. $E_\gamma^{\mathrm{new}} = E_\gamma + \delta_{\mathrm{QRPA}}$, 
so as to match the GDR data
reasonably well around $E_\gamma = 12$ MeV. The resulting theoretical dipole strengths are displayed in 
Figs.~\ref{fig:strength_calc} and~\ref{fig:strength_calc_QRPA} for the best reproduction of
the HFB+c normalization. All the parameters used for the shown models are listed in
Table~\ref{tab:gsfpar}.
\begin{table}[tb]
\begin{center}
\caption{Parameters used for the model strength functions of $^{89}$Y in Figs.~\ref{fig:strength_calc},~\ref{fig:strength_calc_QRPA}.}
\begin{tabular}{lcclcllcc}
\hline
\hline
Nucleus    & $\Gamma_{E1}$ & $E_{E1}$ & $\sigma_{E1}$ & $T_f$ & $\Gamma_{M1}$ & $E_{M1}$ & $\sigma_{M1}$  & $\delta_{\mathrm{QRPA}}$    \\
		   & (MeV)         & (MeV)    &  (mb)         & (MeV) & (MeV)         & (MeV)    & (mb)           & (MeV) \\
\hline
$^{89}$Y   & $4.3$         & $16.8$   & 233.0         & 0.30  & 2.7           & 9.5      & 1.1            & 1.5           \\
\hline
\hline
\end{tabular}
\label{tab:gsfpar}
\end{center}
\end{table}

In the following, we will use our experimentally inferred lower and upper limits 
on the level density and $\gamma$SF as input for cross-section calculations
of the $^{88}$Y($n,\gamma$)$^{89}$Y and $^{88}$Sr($p,\gamma$)$^{89}$Y reactions.

\section{Cross section and rate calculations}
\label{sec:talys}
 \begin{figure*}[!htb]
 \begin{center}
 \includegraphics[clip,width=2\columnwidth]{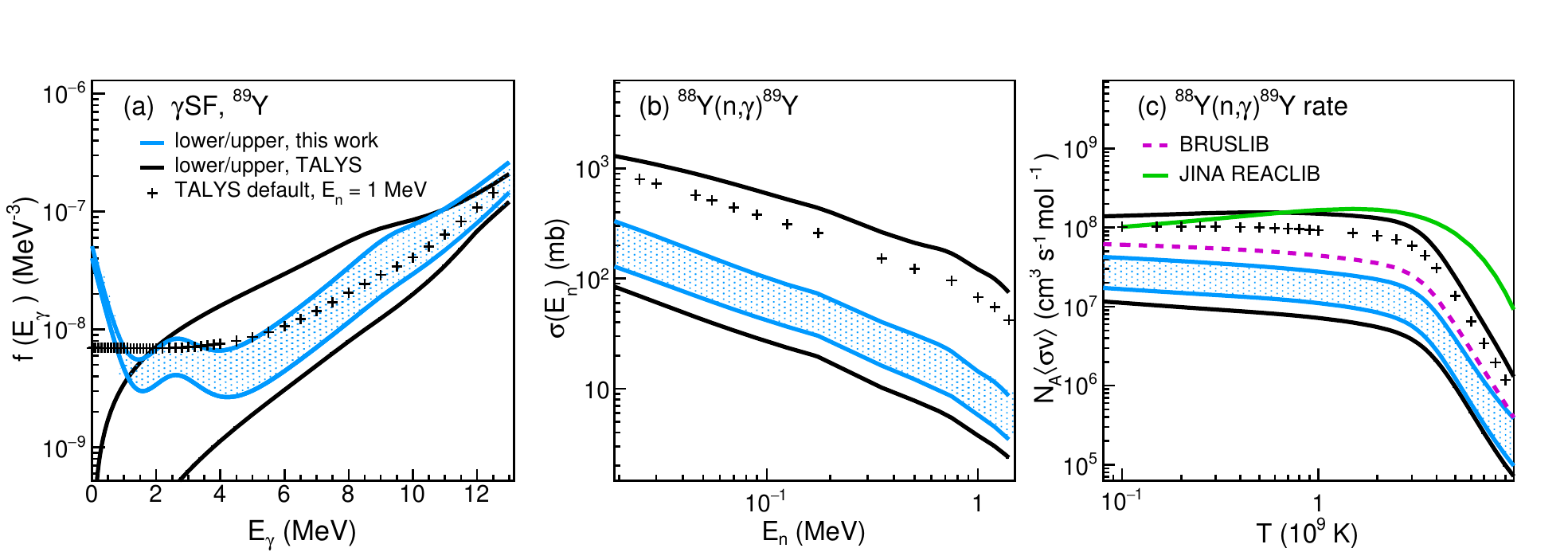}
\caption{(Color online) (a) Input $\gamma$SFs of $^{89}$Y, note that the TALYS default is for $E_{n} = 1$ MeV, 
corresponding to an intial excitation energy of 12.48 MeV in the GLo Model, and that the GLo model is by default normalized to a 
radiative width $\left< \Gamma_\gamma \right> = 170$ meV taken from an interpolation routine in TALYS;
(b) the resulting $^{88}$Y($n,\gamma$)$^{89}$Y cross sections, 
 (c) and the corresponding astrophysical reaction rates compared to the BRUSLIB (dashed magenta line, Ref.~\cite{BRUSLIB}) and the
JINA REACLIB (green solid line, Ref.~\cite{JINA-REACLIB}). The minimum and maximum predictions from the models implemented in TALYS 
    are also shown (thick, black lines).} 
 \label{fig:rates_ng_89Y}
 \end{center}
 \end{figure*}
\begin{table*}[tb]
\begin{center}
\caption{Parameters used for the $\gamma$SF input models in the TALYS calculations.}
\begin{tabular}{lcclcllcccccc}
\hline
\hline
& $\Gamma_{E1}$ & $E_{E1}$ & $\sigma_{E1}$ & $T_f$ & $\Gamma_{M1}$ & $E_{M1}$ & $\sigma_{M1}$  & $\Gamma_{\mathrm{SR}}$ & $E_{\mathrm{SR}}$ & $\sigma_{\mathrm{SR}}$ & $C$                    & $\eta$      \\
& (MeV)         & (MeV)    &  (mb)         & (MeV) & (MeV)         & (MeV)    & (mb)           & (MeV)                  & (MeV)             & (mb)                   & ($10^{-8}$ MeV$^{-3}$) & (MeV$^{-1}$)\\
\hline
Lower limit & $3.5$         & $16.8$   & 233.0         & 0.30  & 2.7           & 9.5      & 0.24           & 2.2                    &  2.8              & 0.10                   & 4.0                    & 2.5     \\
Upper limit & $6.6$         & $17.8$   & 233.0         & 0.30  & 2.7           & 9.5      & 2.00           & 2.2                    &  2.8              & 0.18                   & 5.0                    & 2.5     \\
\hline
\hline
\end{tabular}
\label{tab:talyspar}
\end{center}
\end{table*}

As discussed in the introduction, level density and $\gamma$SF are two key ingredients
in the Hauser-Feshbach approach to calculate cross sections. In this work, we use the open-source 
nuclear reaction code TALYS-1.6~\cite{TALYS_16,koning12} for the cross-section and reaction-rate calculations
for the $^{88}$Sr($p,\gamma$)$^{89}$Y and $^{88}$Y($n,\gamma$)$^{89}$Y reactions.
Our approach is the following:
\begin{itemize}
\item[1.] {Making use of all the various models already implemented in TALYS for the level density, 
the $\gamma$SF, and the proton and neutron
optical potentials to investigate the spread in the resulting cross sections and reaction rates; 
i.e., inferring the lower and
upper limits on these quantities inherent from the available models;}
\item[2.] {Calculating the cross sections and reaction rates with default input parameters in TALYS;}
\item[3.] {Implementing level densities and $\gamma$SFs in accordance 
with our present data and the experimentally inferred 
lower/upper limits.}
\end{itemize}

\textit{1. TALYS predictions for the model uncertainties.\\} 
We have calculated all possible combinations of input level densities, $\gamma$SF models, 
and optical potentials available in TALYS to estimate the minimum and
maximum $^{88}$Y($n,\gamma$)$^{89}$Y and $^{88}$Sr($p,\gamma$)$^{89}$Y cross sections predicted by these models. 
For the ($n,\gamma$) reaction, the combinations are: (\textit{i}) a minimum cross section and rate with 
the temperature-dependent Hartree-Fock-Bogoliubov plus combinatorial
level density of Ref.~\cite{hilaire2012}, the $E1$ strength function from the Hartree-Fock-BCS plus QRPA approach of Ref.~\cite{goriely2002}
making use of the renormalization to the estimated $\left< \Gamma_\gamma\right> = 170$ meV from a spline-fit interpolation table in TALYS,
and the JLM neutron potential~\cite{bauge2001} (TALYS keywords \textit{ldmodel 6, strength 3, gnorm -1., jlmomp y});
(\textit{ii}) a maximum cross section and rate with the combined constant-temperature plus back-shifted Fermi gas model~\cite{gilbert1965}
for the level density with parameters according to the TALYS manual, the standard Lorentzian model (Brink-Axel) 
for the $E1$ strength~\cite{brink1955,axel1962},
and a global parameterization of the neutron optical potential~\cite{koning03} (TALYS keywords \textit{ldmodel1, strength2, localomp n}).
The corresponding combinations for the ($p,\gamma$) reaction are: (\textit{i}) a minimum cross section and rate with 
the temperature-dependent Hartree-Fock-Bogoliubov plus combinatorial
level density of Ref.~\cite{hilaire2012}, the GLo model~\cite{kopecky_uhl_1990} with variable temperature as implemented in TALYS renormalized 
to the spline-fit $\left< \Gamma_\gamma\right>$,
and the global parameterization of the proton optical potential~\cite{koning03} (TALYS keywords \textit{ldmodel 6, strength 1, gnorm -1., localomp n});
(\textit{ii}) a maximum cross section and rate with the combined constant-temperature plus back-shifted Fermi gas model~\cite{gilbert1965}
for the level density with parameters according to the TALYS manual, the standard Lorentzian model (Brink-Axel) 
for the $E1$ strength~\cite{brink1955,axel1962}, and the JLM proton potential~\cite{bauge2001} 
(TALYS keywords \textit{ldmodel 1, strength 2, jlmomp y}).

We find that the level density models in TALYS give a factor of $\approx 5$ and $\approx 4$ uncertainty 
for the ($n,\gamma$) and ($p,\gamma$)
cross sections, respectively, 
while the corresponding numbers for the $\gamma$SF models are $\approx 30$ and $\approx 28$, respectively.
The impact of the choice of optical-model potential has also been tested for the reactions of interest.
More specifically, we have used the proton and neutron potentials of
Koning and Delaroche~\cite{koning03} with global parameters as described
in the TALYS manual, and also the semi-microscopic optical potential of the Jeukenne-Lejeune-Mahaux (JLM) type~\cite{bauge2001};
see the TALYS documentation for more details~\cite{TALYS_16,koning12}. 
We did not adjust any parameters in the neutron or proton potentials, but used the default parameters as implemented
in TALYS-1.6. We have found that the ratio between the cross sections using the global potential and the JLM potential reaches 
a maximum deviation of $\approx 34$\% and
$\approx 48$\% for the proton and neutron potentials, respectively, for the energy ranges $2.5 \cdot 10^{-6} \leq E_n \leq 5$ MeV and
$0.75 \leq E_p \leq 5$ MeV. On average, the JLM potential gives a lower capture cross section than the global potential for the neutron capture,
and opposite for proton capture.
\begin{figure*}[!htb]
\begin{center}
\includegraphics[clip,width=1.4\columnwidth]{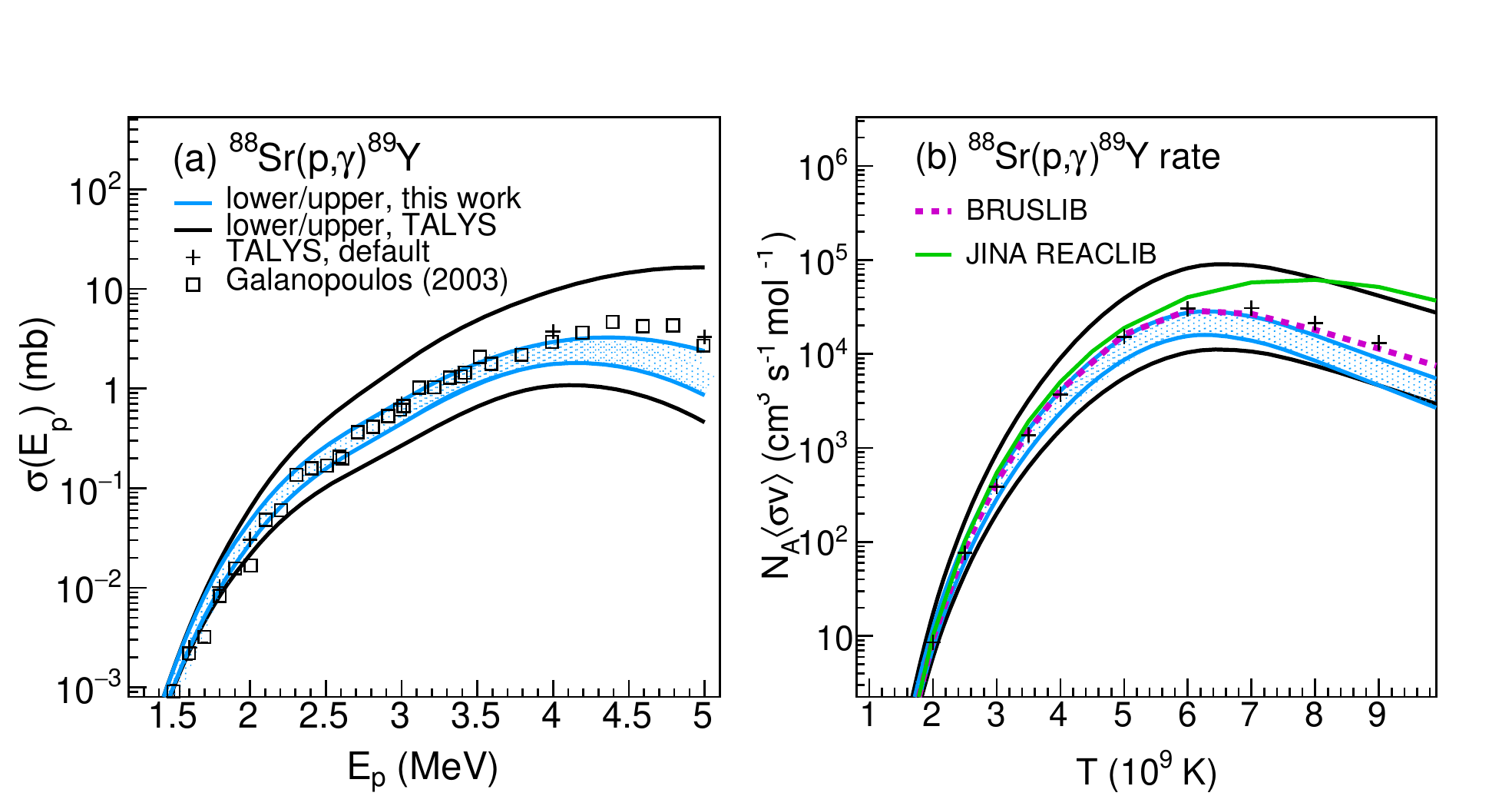}
\caption{(Color online) Calculated $^{88}$Sr($p,\gamma$)$^{89}$Y cross sections shown as a blue-shaded band 
(a) compared to data from Ref.~\cite{galanopoulos2003}, 
and the corresponding astrophysical reaction rates (b) compared to the BRUSLIB (dashed magenta line, Ref.~\cite{BRUSLIB}) and the
JINA REACLIB (green solid line, Ref.~\cite{JINA-REACLIB}). The minimum and maximum predictions from the models implemented in TALYS 
are also shown (thick, black lines), as well as the result using default input parameters (black crosses).} 
\label{fig:rates_pg_89Y}
\end{center}
\end{figure*}

The combined model uncertainties in the cross sections, including the optical-model uncertainties, reach a factor of $\approx 33$ and 
$\approx 36$ at maximum for the ($n,\gamma$)$^{89}$Y and ($p,\gamma$)$^{89}$Y cross sections.
These can be considered as the intrinsic 
uncertainties for the models implemented in TALYS (note again that no information is available regarding the level spacing $D_0$ and the
average radiative width $\left< \Gamma_\gamma \right>$). The results are shown as thick, black lines in 
Figs.~\ref{fig:rates_ng_89Y} and~\ref{fig:rates_pg_89Y}. 

\textit{2. TALYS default predictions.\\}
If no keywords related to level density, $\gamma$SF, or proton or neutron potential are specified in the TALYS input file,
default values will automatically be invoked. For the ($n,\gamma$)$^{89}$Y reaction, these are 
the neutron optical potential~\cite{koning03} with local parameters, the GLo model with variable temperature~\cite{kopecky_uhl_1990},
and  the combined constant-temperature plus back-shifted Fermi gas model~\cite{gilbert1965} for the level density. 
For the ($p,\gamma$)$^{89}$Y reaction these default values are the same, except for the proton potential, for which there are no local
parameters and the global parameterization of Ref.~\cite{koning03} is used. The results are displayed as black crosses in 
Figs.~\ref{fig:rates_ng_89Y} and~\ref{fig:rates_pg_89Y}.

It is interesting to note that, by default, there is an automatic re-scaling of the $\gamma$SF in TALYS to match the
estimated  $\left< \Gamma_\gamma\right> = 170$ meV from the spline-fit interpolation table. However, this procedure does 
not guarantee that the resulting $(n,\gamma)$ cross section will be the same for different model combinations of the 
$\gamma$SF and level density. The default-estimate cross section (crosses in
Fig.~\ref{fig:rates_ng_89Y}) 
with a scaling factor \textit{gnorm} = 1.44 is very different from the lower-limit cross section 
with \textit{gnorm} = 0.28 (lower, thick black lines
in Fig.~\ref{fig:rates_ng_89Y}). This demonstrates that there is a delicate interplay with the adopted level density 
(with its spin distribution), $\gamma$SF, and particle optical potential in the calculation of radiative cross sections.

\textit{3. Results from this work implemented in TALYS.\\}
Finally, we have used our results to constrain the input level density and $\gamma$SF for the
$^{88}$Y($n,\gamma$)$^{89}$Y and $^{88}$Sr($p,\gamma$)$^{89}$Y cross sections and reaction rates. 
We have used the constant-temperature model for the level density with parameters given in Sec.~\ref{sec:nld}
above $E_x = 2.88$ MeV, with a spin distribution and parameters according to the lower and upper normalizations 
as described in Sec.~\ref{sec:nld}. This model reproduces our level-density data very well.
Below that excitation energy, we use the known, discrete levels.
For the $\gamma$SF data, we have tuned the $\gamma$SF models to reproduce our lower and upper
data points. Specifically, we have used a low-lying $M1$ strength that corresponds to the shell-model
results parameterized as an exponential function $f_{\mathrm{upbend}} = C\exp{-\eta E_\gamma}$, 
an $M1$ spin-flip resonance, and a GLo $E1$ component so as to match our upper and lower limits.
In addition, our $\gamma$SF data points overshoot the $M1+E1$ models in the energy range of $E_\gamma \approx 2.0-3.5$ MeV,
as seen in Figs.~\ref{fig:strength_calc} and~\ref{fig:strength_calc_QRPA}. Therefore, 
we have added a small resonance (SR) of $M1$ type to get a reasonable
agreement with the measured strength. The applied parameters are given in Tab.~\ref{tab:talyspar}. 
The input $\gamma$SF for $^{89}$Y is shown in Fig.~\ref{fig:rates_ng_89Y}(a).

The resulting $^{88}$Y($n,\gamma$)$^{89}$Y and $^{88}$Sr($p,\gamma$)$^{89}$Y
cross sections are displayed in Figs.~\ref{fig:rates_ng_89Y}(b),~\ref{fig:rates_pg_89Y}(a)
respectively, and shown as blue-shaded bands. 
We note that our estimated ($n,\gamma$) cross section is in the lower region of the uncertainty
band provided by TALYS. Also, our error band does not match well with the default TALYS prediction, 
as our results indicate a significantly lower cross section.

For the $^{88}$Sr($p,\gamma$)$^{89}$Y cross section,
direct measurements exist~\cite{galanopoulos2003}. Overall, we find a very good agreement
between our calculations and the $^{88}$Sr($p,\gamma$)$^{89}$Y data from Ref.~\cite{galanopoulos2003}, although the 
data indicate a slightly higher cross section for incoming proton energies $E_p > 4.3$ MeV.
The overall good agreement, however, gives confidence in our approach to estimate radiative capture 
cross sections, as also demonstrated in Refs.~\cite{toft2011,tornyi2014}.

We have also estimated the astrophysical reaction rates, which are shown in 
Figs.~\ref{fig:rates_ng_89Y}(c),~\ref{fig:rates_pg_89Y}(b). We have compared to the BRUSLIB~\cite{BRUSLIB} and the
JINA REACLIB~\cite{JINA-REACLIB} databases. We find that the BRUSLIB rates are rather similar to
our estimated upper limits, especially for the ($p,\gamma$) reaction. 
In contrast, the JINA REACLIB rates are significantly higher both for the proton-capture and
neutron-capture cases, especially for higher stellar temperatures.

The low-energy upbend has been shown to increase neutron-capture rates of very neutron-rich nuclei, if it is found to be present in
such exotic isotopes~\cite{larsen2010}. For $^{89}$Y, however, the impact of the upbend on the cross sections is
very limited; for example, for the ($p,\gamma$) reaction, the relative change in cross section is at maximum 1\%, 
and for the $^{88}$Y($n,\gamma$) reaction maximum 2\%. However, provided that the upbend
persists for very neutron-rich Y nuclei, it may influence the ($n,\gamma$) rates of relevance for the $r$-process.

\section{Summary and outlook}
\label{sec:sum}
In this work, we have determined the nuclear level density and $\gamma$SF of $^{89}$Y. 
The data are available at the following web site: 
\url{http://ocl.uio.no/compilation}.
We find that this nucleus exhibits a
low-energy enhancement in the $\gamma$SF, in accordance with previous findings in this mass region. Moreover, 
shell model calculations describe the main part of the low-energy 
enhancement by $M1$ transitions, similar to previous shell-model results 
for Mo, Fe, and Zr isotopes. However, as (a contribution of) $E1$ transitions cannot be ruled out based on the available
data, an experimental determination of the electromagnetic character of this enhancement
is highly desired to firmly establish the mechanism behind this phenomenon.

We have applied our new data as input for nuclear-reaction calculations, estimating radiative capture cross sections
and reaction rates for the reactions $^{88}$Sr($p,\gamma$)$^{89}$Y and  $^{88}$Y($n,\gamma$)$^{89}$Y.
For the radiative proton capture case, we find good agreement with our experiment-constrained calculations and
direct measurements. Moreover,
in comparison with widely used reaction-rate libraries, we note that the BRUSLIB predictions in general agree better 
with our results. 

In the future, it would be very valuable to measure other nuclei in this mass region with the
Oslo method, to be able to further constrain reactions such as $^{87}$Y($n,\gamma$)$^{88}$Y of importance to
the $p$-process nucleosynthesis. To be able to reduce the errors associated with our approach, 
further developments of independent absolute-normalization techniques would be highly desirable, 
as well as complementary experiments, such as measurements of particle-evaporation spectra and
$\gamma$-ray two-step cascade spectra to infer level densities and $\gamma$SFs, respectively.

\acknowledgments

The authors wish to thank J.C.~M{\"{u}}ller, E.A.~Olsen, A.~Semchenkov and J.~Wikne at the 
Oslo Cyclotron Laboratory for providing excellent experimental conditions. 
This work was financed in part by the Research Council of Norway (NFR), project grant no. 205528, and 
through the ERC-STG-2014 under grant agreement no. 637686. S.~G. is FNRS research associate.
S.~S. acknowledges financial support
by the NFR under project grant no. 210007.
M.~W. acknowledges support by the National Research Foundation of South Africa under grant no. 92789.

\vfill
\end{document}